# A Statistical Analysis of Recent Traffic Crashes in Massachusetts

Aaron Zhang

*Acton-Boxborough Regional High School*
Acton, MA 01720, USA

aaronhxzhang@gmail.com

Evan W. Patton

*CSAIL, Massachusetts Institute of Technology*
Cambridge, MA 02139, USA

ewpatton@mit.edu

Justin M. Swaney*

*Department of Chemical Engineering*
*Massachusetts Institute of Technology*
Cambridge, MA 02139, USA

jswaney@mit.edu

Tingying Helen Zeng*

Division of Career Education
Academy for Advanced Research and Development
14[th] floor, One Broadway, Cambridge, MA 02142, U.S.A.

helen.zeng@us-aard.com

*Co-corresponding authors.

*Abstract* – **A statistical analysis implemented in the Python programming language was performed on the available MassDOT car accident data to identify whether a certain set of traffic circumstances would increase the likelihood of injuries. In the analysis, we created a binary classifier as a model to separate crashes that resulted in injury from those that did not. To accomplish this, we first cleaned up the initial data, then proceeded to represent categorical variables numerically through one hot encoding before finally producing models with *Recursive Feature Elimination (RFE)* and without RFE, in conjunction with logistic regression. This statistical analysis plays a significant role in our modern road network that has presented us with a heap of obstacles, one of the most critical being the issue of how we can ensure the safety of all drivers and passengers. Findings from our analysis identify that tough weather and road conditions, senior/teen drivers and dangerous intersections play prominent roles in accidents that resulted in injuries in Massachusetts. These new findings can provide valuable references and scientific data support to relevant authorities and policy makers for upgrading road infrastructure, passing regulations, etc.**

## I. INTRODUCTION

A quick glance of some fast statistics immediately reveals the severity of the safety issue our roads are currently facing. In the United States alone, a staggering 6 million car accidents occur each year which equates to over 16,000 crashes every single day. [1] According to the data gathered by the World Health Organization (WHO): About 1.35 million people in the world die from road traffic crashes each year. 20 to 50 million more people suffer non-fatal injuries, with many resulting in disability. "Road traffic injuries cause considerable economic losses to individuals, their families, and to nations as a whole. These losses arise from the cost of treatment as well as lost productivity for those killed or disabled by their injuries, and for family members who need to take time off work or school to care for the injured." [2]

Today traffic accident data is widely analyzed across the world and progress is being made to improve road safety and to decrease the amount of traffic accidents involving injuries. As an example, some recent data analysis was performed to determine high-frequency accident locations. [3] The objective of our work is to determine the factors that have the greatest capacity to cause serious injury and subsequently recommend changes to address these factors in Massachusetts. For instance, after finding a certain type of intersection to correlate strongly with crashes involving injuries, we have the ability to monitor these crashes with the progression of time. Our goal, in that scenario, would therefore be to eventually make infrastructure recommendations to town, state, and federal governments.

## II. MATERIALS AND METHOD (PROCESS)

The purpose of this study is to determine the greatest risk factors for crashes that result in injury such that a greater emphasis can be placed upon eliminating those risks on our roads in Massachusetts. To determine these factors most strongly associated with traffic accidents today, we created a binary classifier as a model to separate crashes resulted in injuries from crashes that did not through the following procedure: first, we obtained a random dataset of crashes in Massachusetts within the past 5 years, where every entry contained specific categories specifying the details of a crash, and cleaned up the dataset by removing all entries containing null values to prevent any errors during coding. We then proceeded to represent categorical variables numerically through one hot encoding/number encoding before finally producing a model through logistic regression to identify whether a certain set of circumstances will increase the likelihood of injuries. The process is illustrated in the diagram in Figure 1.

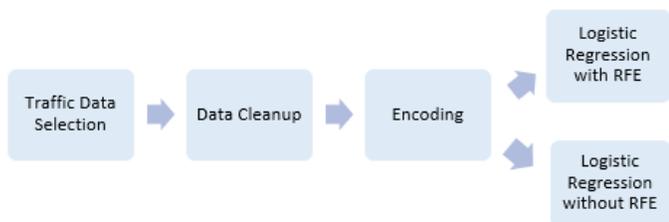

Figure 1: Traffic data analysis process

Our statistical analysis is implemented in Python. We utilized Python packages including scikit-learn, pandas and NumPy. The original code used for this analysis is available at https://github.com/aaron1116/Data-Science-in-Traffic-Incident-Analysis.

*A. Dataset Selection*

The dataset for this paper was a randomly selected sample of 46,077 crash entries for the last five years retrieved from MassDOT's crash portal [4] where every traffic accident in Massachusetts is listed with a multitude of attributes including, but not limited to: number of injuries, manner of collision, ages of drivers, weather condition during the accident, etc. This dataset was selected because of the abundance of data provided, allowing for a much more thorough analysis.

Although the dataset does offer plenty of details, the greatest issue with the dataset is its lack of numerical data which posed a significant challenge later. The only categories containing numerical data were speed limit, Linked Road Inventory Average Daily Traffic, and Linked Road Inventory Number of Travel Lanes.

Figure 2: MassDOT Crash Dataset (https://www.mass.gov/crash-data, accessed during 06/24/2019 – 09/20/2019 [5])

*B. Cleaning Up the Dataset*

A slight problem with the dataset was the excessive amount of data, specifically that many of the categories could potentially become distracting to the final results of the logistic regression. To resolve this issue, we removed columns such as coordinates of the crash and address of the crash which have too many distinct possibilities to group together.

*C. One Hot Encoding/Number Encoding*

The most critical step was to represent all of the categorical data numerically. To accomplish this, we utilized the technique of one hot encoding/number encoding. One hot encoding/number encoding is a method by which categorical variables are converted into "numerical" variables through "encoding" each categorical value as either an array of 0s and 1s or more simply, a distinct integer. [6] This technique was particularly crucial due to the lack of numerical data which was mentioned in section A. Furthermore, the encoding of categorical variables was paramount for each of those variables to be accounted for as an input independent variable in the logistic regression model.

*D. Logistic Regression*

Logistic regression is a statistical method to model a "binary" dependent variable (i.e. has two possible values) by assigning weights to the independent explanatory variables and subsequently using those weights to predict between the two possible outcomes of the dependent variable. [6]

Our objective was to use logistic regression to create a model to identify whether a certain set of conditions would increase the likelihood of an injury-inducing accident, then extracting the weights the model employed to determine the factors most associated with more serious crashes.

*E. Recursive Feature Elimination (RFE)*

The Recursive Feature Elimination (RFE) method is a feature selection approach. It works by recursively removing attributes and building a model on those attributes that remain. It uses the model accuracy to identify which attributes (and combination of attributes) contribute the most to predicting the target attribute. [7]

Initially, we attempted to use Recursive Feature Elimination (RFE) in conjunction with logistic regression to determine the variables which had the greatest and least impact when predicting whether crashes would result in injury or not.

### III. RESULTS AND DISCUSSION

To compare the significance of each factor in relation to the model as a whole, we listed out the weights of each variable as determined by the model in ascending order. A bar graph for ease of interpretation is included as well. In addition, we also provided the breakdown for the top factors that contributed to crashes involving injuries in order to create a better understanding of those factors.

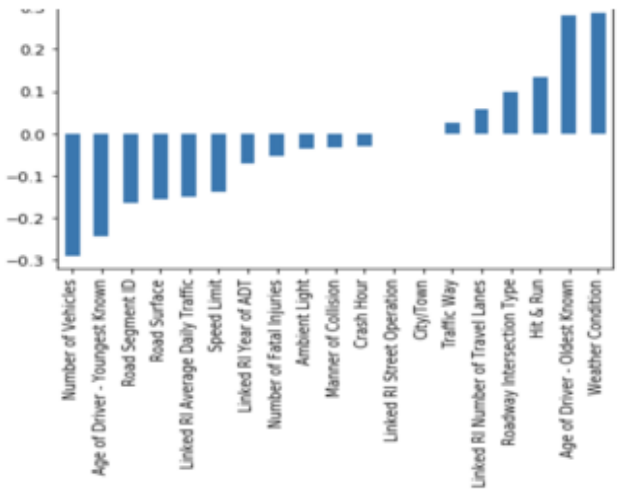

```
Weather Condition                        0.286713
Age of Driver - Oldest Known             0.280260
Hit & Run                                0.133549
Roadway Intersection Type                0.099435
Linked RI Number of Travel Lanes         0.058910
Traffic Way                              0.026487
City/Town                                0.000000
Linked RI Street Operation               0.000000
Crash Hour                              -0.031238
Manner of Collision                     -0.031818
Ambient Light                           -0.036903
Number of Fatal Injuries                -0.052769
Linked RI Year of ADT                   -0.070865
Speed Limit                             -0.137935
Linked RI Average Daily Traffic         -0.149021
Road Surface                            -0.154812
Road Segment ID                         -0.163574
Age of Driver - Youngest Known          -0.243217
Number of Vehicles                      -0.291274
dtype: float64
```

Figure 3: Variables used for the logistic regression model ordered by weight

| Top factors that contributed to crashes involving injuries | Breakdown of each factor | Weights |
|---|---|---|
| **Weather Condition** | | |
| | Snow/sleet, hail (freezing rain or drizzle) | 0.590855 |
| | Rain/Rain | 0.439613 |
| | Snow/Snow | 0.335264 |
| | Snow/Blowing sand, snow | 0.295555 |
| | Cloudy/Clear | 0.254933 |
| | Snow/Cloudy | 0.253511 |
| | Nor reported | 0.243501 |
| | Rain/Snow | 0.104463 |
| | Cloudy/cloudy | -0.105724 |
| **Road Surface** | | |
| | Ice | 0.506753 |
| | Slush | 0.260811 |
| | Not reported | 0.243501 |
| | Water (standing, moving) | 0.166821 |
| **Age of Driver** | | |
| | Oldest known_75-84 | 0.465152 |
| | Youngest known_16-20 | 0.322298 |
| | Youngest known_65-74 | 0.127612 |
| | Oldest known_25-34 | 0.123682 |
| | Youngest known_21-24 | -0.106325 |
| **Roadway Intersection Type** | | |
| | Y-intersection | 0.182016 |
| | Driveway | 0.163323 |
| | Five-point or more | 0.141659 |

Figure 4: Breakdown of top factors contributed to crashes involving injuries

Figure 3 depicts the variables which contributed to the logistic regression model in ascending order by weight, while Figure 4 shows the breakdown of the top factors that contributed to the crashes involving injuries.

According to Figure 3 and Figure 4, the factors that appear to be contribute the greatest weight to the model are tough weather and road conditions, senior/teen drivers (Age 75-84 and Age 16-20), and dangerous intersections. This ranking is logical given the challenging weather conditions that Massachusetts notoriously features. We can potentially generate very different rankings of the factors by performing the same statistical analysis on crash data from different states across the United States.

Based on our statistical analysis results, to effectively reduce the number of injury-inducing traffic accidents in Massachusetts, we would highly recommend relevant authorities and policy makers to prioritize investing in solutions including improving weather forecast accuracy and implementing traffic bans when needed, investing further in regular road surface maintenance in addition to increasing efforts to clear roads during times of severe precipitation, improving regulations to protect senior and teen drivers, and updating road designs to avoid construction of intersections that correlate with high risk of traffic accidents resulting in injuries.

As mentioned previously, we created two models to compare accuracy: one with RFE and one without. The model without RFE achieved a slightly higher training accuracy of 75.5%, while the model with RFE that included 10 out of 21 available variables achieved 73%. The most likely reason for the slight discrepancy in accuracy is that RFE limits the number of features available for logistic regression by removing some features with slight predictive power.

For future efforts in forecasting injuries that result from traffic accidents, we should construct a more predictive model and consider ways of further increasing the model accuracy. As mentioned previously, though crash data is better interpreted by linear modelling, it may be particularly tough to separate crash data by a linear classifier. The most logical way to proceed is to thus to create a non-linear classifier instead of using logistic regression. We can implement a technique such as Support Vector Machine (SVM) using the "kernel trick". In machine learning, SVMs are supervised learning models with associated learning algorithms that analyze data used for classification and regression analysis. In addition to performing linear classification, SVMs can efficiently perform a non-linear classification. [8] In place of a model which creates a best fit line, the SVM will create a "best fit hyperplane" in multi-dimensional space, then apply it in a low-dimensional space as a "non-linear" classifier of the data.

## IV. CONCLUSION

Based on the available MassDOT data of car accidents from the past five years, we proposed using logistic regression in conjunction with Recursive Feature Elimination (RFE) and one hot encoding/number encoding. We created models with and without RFE, then determined the weights of different factors that the models consisted of using Python. Our results show that the greatest potential factors of serious accidents involving injuries include tough weather and road conditions, senior/teen drivers (Age 75-84 and Age 16-20) and dangerous intersections. These new findings provide scientific support to motivate changes in our approach to traffic safety with more emphasis being placed on addressing these factors with higher priority. For instance, additional safety precautions or policies should be considered to protect senior/teen drivers on Massachusetts's roads or to avoid construction of intersections that correlate with high rates of serious crashes.

Our model is adaptable towards accident data outside of Massachusetts. In fact, our goal is to eventually create models for crashes across the United States via a federal database of crashes. We believe when this is achieved, our data analysis results will provide an even greater impact and benefit more people.

ACKNOWLEDGMENTS

We are very grateful for the sponsorship and internship training by the Academy for Advanced Research and Development (AARD) during the development of this project. The AARD website is: http://www.us-aard.com. The first author is grateful for the supervision of Justin M. Swaney during the summer research program and is thankful to Dr. Evan Patton for his thoughtful comments that helped guide this research.

REFERENCES

[1] DriverKnowledge. (2019). *Car Accident Statistics in the U.S. | Driver Knowledge*. [online] Available at: https://www.driverknowledge.com/car-accident-statistics
[2] World Health Organization (WHO), "Road traffic injuries", https://www.who.int/en/news-room/fact-sheets/detail/road-traffic-injuries
[3] Chen, Chen, "Analysis and Forecast of Traffic Accident Big Data", https://www.researchgate.net/publication/319487117_Analysis_and_Forecast_of_Traffic_Accident_Big_Data
[4] MassDOT. (2019). *Crash Query and Visualization*. [online] Available at: https://apps.impact.dot.state.ma.us/cdv
[5] MassDOT Crash Dataset, https://www.mass.gov/crash-data, accessed during 06/24/2019 – 09/20/2019
[6] Yin, Julie. "Use One-Hot-Encoding To Analyze Adult Income Data", Medium, https://medium.com/@julie.yin/use-one-hot-encoding-to-analyze-adult-income-data-and-some-bad-news-for-the-single-people-in-the-cef71f9d47b4
[7] Bakharia, Aneesha. "Recursive Feature Elimination with Scikit Learn", Medium, https://medium.com/@aneesha/recursive-feature-elimination-with-scikit-learn-3a2cbdf23fb7
[8] Patel, Savan. "SVM (Support Vector Machine) — Theory", Medium, https://medium.com/machine-learning-101/chapter-2-svm-support-vector-machine-theory-f0812effc72